\documentclass[superscriptaddress,prl,twocolumn,nofootinbib,showpacs,preprintnumbers]{revtex4}

\usepackage{amsfonts}
\usepackage{mathrsfs}
\usepackage{amssymb}
\usepackage{amsmath}
\usepackage{graphicx,epsfig}

\def\bwt{\begin{widetext}}

\def\ewt{\end{widetext}}

\def\be{\begin{equation}}

\def\ee{\end{equation}}

\def\bea{\begin{eqnarray}}

\def\eea{\end{eqnarray}}

\def\bean{\begin{eqnarray*}}

\def\eean{\end{eqnarray*}}

\def\bary{\begin{array}}

\def\eary{\end{array}}

\def\bit{\begin{itemize}}

\def\eit{\end{itemize}}

\def\su5u1{SU(5) \times U(1)}

\def\fsu5u1{SU(5) \times U(1)'}

\def\so10{SO(10)}

\def\sq20{SO(10) \times SO(10)}

\usepackage{amssymb}

\begin{document}

\setlength{\parskip}{0cm}

\title{F-Theory Grand Unification at the Colliders }

\author{Tianjun Li}

\affiliation{George P. and Cynthia W. Mitchell Institute for
Fundamental Physics, Texas A$\&$M University, College Station, TX
77843, USA }

\affiliation{Key Laboratory of Frontiers in Theoretical Physics,
      Institute of Theoretical Physics, Chinese Academy of Sciences,
Beijing 100190, P. R. China }

\author{James A. Maxin}

\affiliation{George P. and Cynthia W. Mitchell Institute for
Fundamental Physics, Texas A$\&$M University, College Station, TX
77843, USA }

\author{Dimitri V. Nanopoulos}

\affiliation{George P. and Cynthia W. Mitchell Institute for
Fundamental Physics,
 Texas A$\&$M University, College Station, TX 77843, USA }

\affiliation{Astroparticle Physics Group,
Houston Advanced Research Center (HARC),
Mitchell Campus, Woodlands, TX 77381, USA}

\affiliation{Academy of Athens, Division of Natural Sciences,
 28 Panepistimiou Avenue, Athens 10679, Greece }

%\date{\today}

%%%%%%%%%%%%%%%%%%%%%%%%%%%%%%%%%%%%%%%%%%%%%%%%%%%%%%%%%%%%%%%%%%%%%%%%%%%%

\begin{abstract}

We predict the exact gaugino mass relation near 
the electroweak scale at one loop for gravity 
mediated supersymmetry breaking in F-theory 
$SU(5)$ and $SO(10)$ models with  $U(1)_Y$ 
and $U(1)_{B-L}$ fluxes, respectively. The 
gaugino mass relation introduced here differs 
from the typical gaugino mass relations studied 
thus far, and in general, should be preserved
quite well at low energy. Therefore, these F-Theory 
models can be tested at the Large Hadron Collider 
and future International Linear Collider. 
We present two typical scenarios that satisfy 
all the latest experimental constraints and are 
consistent with the CDMS II experiment. In 
particular, the gaugino mass relation is 
indeed satisfied at two-loop level 
with only a very small deviation around the 
electroweak scale.

\end{abstract}


\preprint{ACT-01-10, MIFP-10-03}

\maketitle

\section{Introduction}

The great challenge of string phenomenology
is constructing realistic string models which allow us to make unique 
predictions that can be tested at the Large Hadron Collider (LHC), 
future International Linear Collider (ILC), and other
experiments. If these predictions are confirmed at
future experiments, we will possess 
strong evidence to support that string theory is indeed the correct 
fundamental description of nature. 
To the present, string phenomenology has been primarily 
concentrated on heterotic $E_8\times E_8$ string theory and 
Type II string theories with D-branes, though unfortunately, 
this has not resulted in any unique
prediction thus far.

The last two years have seen Grand Unified Theories (GUTs)
constructed locally in F-theory, which can be considered as the
strongly coupled formulation of ten-dimensional Type IIB string 
theory with a varying axion-dilaton field 
$S$~\cite{Vafa:1996xn, Donagi:2008ca, Beasley:2008dc, 
Beasley:2008kw, Donagi:2008kj}. In F-theory model building, 
the gauge fields reside on the observable seven-branes that wrap a del Pezzo 
$n$ ($dP_n$) surface for the extra four space dimensions, 
while the  Standard Model (SM) fermions
and Higgs fields are localized on the complex
codimension one curves (two-dimensional real subspaces) in $dP_n$. 
Certainly, F-Theory model building and phenomenology have been
studied extensively~\cite{Font:2008id, Jiang:2009zza, 
Blumenhagen:2008aw, Heckman:2009bi, Donagi:2009ra, 
Jiang:2009za, Li:2009cy, Li:2009fq}.
In contrast to D-brane model building~\cite{Berenstein:2006aj}, 
all the SM fermion Yukawa couplings 
can be obtained from the triple
intersections of the SM fermion and Higgs curves.
An exciting new feature is that $SU(5)$ gauge symmetry
can be broken down to the SM gauge symmetry
by turning on $U(1)_Y$ 
flux~\cite{Beasley:2008dc, Beasley:2008kw, Li:2009cy}, 
and additionally, the $SO(10)$ gauge 
symmetry can be broken down to the $SU(5)\times U(1)_X$
and $SU(3)\times SU(2)_L\times SU(2)_R\times U(1)_{B-L}$
gauge symmetries by turning on the $U(1)_X$ and $U(1)_{B-L}$
fluxes, respectively~\cite{Beasley:2008dc, Beasley:2008kw, 
Jiang:2009zza, Jiang:2009za, Font:2008id, Li:2009cy}. 
It is significant to note that realistic GUTs from F-theory can be constructed
locally, hence, the next key question is whether a unique
prediction can be made that can tested at 
the LHC, ILC, and other future experiments.

To study the low energy phenomenology from F-theory GUTs, 
gauge mediated supersymmetry breaking 
was predominantly considered since the F-theory GUTs were constructed 
locally~\cite{Heckman:2009bi}. However, to construct realistic F-theory GUTs,
we must embed these local GUTs into a globally consistent 
framework~\cite{Donagi:2009ra}.
Consequently, here we study gravity mediated supersymmetry breaking. In F-theory  
$SU(5)$ and  $SO(10)$ models where the gauge symmetries are 
broken down
to the  $SU(3)\times SU(2)_L\times U(1)_Y$ and 
 $SU(3)\times SU(2)_L\times SU(2)_R\times U(1)_{B-L}$
gauge symmetries by turning on the $U(1)_Y$ and $U(1)_{B-L}$
fluxes, respectively, we obtain the exact
gaugino mass relation
(See Eq.~(\ref{GMRelation}) in the following)
near the electroweak scale at one loop.
These F-theory GUTs are constructed locally, 
so we do not know the K\"ahler potential for the SM fermions 
and Higgs fields. For this reason, we cannot calculate the
supersymmetry breaking
scalar masses and trilinear soft terms.
Interestingly, our gaugino mass relation can be preserved
very well at the low energy two-loop level if the 
scalar masses and trilinear soft terms
are near the TeV scale. Using the indices for the gaugino mass
relations~\cite{Li:2010xr, Li:2010hi}, we show that
our gaugino mass relation is different from those that have been
studied thus far~\cite{Choi:2007ka}. Because the gaugino masses can be measured
at LHC and ILC~\cite{Cho:2007qv, Barger:1999tn, Altunkaynak:2009tg},
these F-theory GUTs can be tested at the colliders.
Note that the generic scalar masses and trilinear soft terms will not affect our
prediction on the gaugino mass relation at low energy,
so we assume a universal scalar mass $m_0$ and universal trilinear soft 
term $A_{0}$ for simplicity. Examining two typical scenarios of 
gaugino masses, we present the viable parameter space which
satisfies all the latest experimental constraints and is consistent with
the CDMS II experiment~\cite{Ahmed:2009zw}. In particular, the gaugino mass 
relation is in fact satisfied at two-loop level with only a very slight
deviation at low energy. More detailed discussions will be
presented elsewhere~\cite{LMN-Preparation}.

\section{Gaugino Mass Relation}

In the F-theory GUTs, the  GUT gauge symmetries on
the observable seven-branes are broken by turning on
the $U(1)$ fluxes. Interestingly,  these $U(1)$ fluxes will
give additional 
contribution to the gauge kinetic functions, which can be computed 
by dimensionally reducing the Chern-Simons action of the
observable seven-branes wrapping on $dP_n$
\begin{eqnarray}
S_{\rm CS} &=& \mu_7 \int_{dP_n\times \mathbb{R}^{3,1}} a \wedge {\rm tr}(F^4) ~.~\,
\end{eqnarray}
For simplicity, we will assume that the heavy 
KK states and string states have masses above the GUT scale,
which can be realize naturally in the global F-theory GUTs.

First, let us consider the $SU(5)$ 
models~\cite{Beasley:2008dc, Beasley:2008kw, Li:2009cy}. 
Turning on the $U(1)_Y$ flux, the gauge kinetic functions
$f_3$, $f_2$ and $f_1$ respectively
for $SU(3)_C$, $SU(2)_L$ and $U(1)_Y$ gauge symmetries
at the string scale can be
parametrized as follows~\cite{Donagi:2008kj, Blumenhagen:2008aw}
\begin{eqnarray}
f_3 &=& \tau + {1 \over 2} \alpha S~,~\,  
f_2 ~=~ \tau + {1 \over 2} \left(\alpha+2 \right) S~,~ \nonumber \\
f_1 &=& \tau + {1 \over 2} \left(\alpha+{6\over 5}  \right) S~,~\,
\end{eqnarray}
where $\tau$ is the original gauge kinetic function of
$SU(5)$, the $S$ terms arise from $U(1)_Y$ flux contributions,
and $\alpha$ is a positive integer.

Second, let us consider the $SO(10)$ models. If the $SO(10)$ gauge symmetry
is broken down to the flipped $SU(5)\times U(1)_X$ gauge symmetry
by turning on the $U(1)_X$ flux~\cite{Beasley:2008dc, Beasley:2008kw,
Jiang:2009zza, Jiang:2009za}, we can show that the gauge kinetic
functions for $SU(5)$ and $U(1)_X$ are exactly the 
same at the unification scale~\cite{Jiang:2009za}. Interestingly,
if we break the $SO(10)$ gauge symmetry down to the 
$SU(3)\times SU(2)_L\times SU(2)_R\times U(1)_{B-L}$
gauge symmetry by turning on the $U(1)_{B-L}$ 
flux~\cite{Font:2008id, Li:2009cy},
we can show that the gauge kinetic functions 
for the $SU(3)_C$, $U(1)_{B-L}$, $SU(2)_L$, and $SU(2)_R$
gauge symmetries at the string scale are~\cite{Li:2009cy}
\begin{eqnarray}
f_{SU(3)_C} &=& f_{U(1)_{B-L}}= \tau + S ~,~\, \nonumber \\
f_{SU(2)_L} &=& f_{SU(2)_R} = \tau ~,~\,
\end{eqnarray}
where $\tau$ is the original gauge kinetic function of $SO(10)$,
and the $S$ term arises from $U(1)_{B-L}$ flux contribution.
We can break the $SU(2)_R\times U(1)_{B-L}$ gauge symmetry
down to $U(1)_Y$ at the string scale by the Higgs mechanism. 
As a consequence, we obtain the gauge kinetic function
for $U(1)_Y$~\cite{Li:2009cy}  
\begin{eqnarray}
 f_{U(1)_{Y}}= {3\over 5} f_{SU(2)_R} + {2\over 5} f_{U(1)_{B-L}}=
\tau + {2\over 5} S ~.~\,
\end{eqnarray}

Now, let us study gravity mediated supersymmetry breaking.
We can show that the gaugino mass relation in 
the $SO(10)$ models with $U(1)_{B-L}$ flux
is the same as that in the $SU(5)$ models with $U(1)_Y$ flux.
Henceforth, we only consider 
the $SU(5)$ models with $U(1)_Y$ flux here.
Supposing supersymmetry is broken by the F-terms of $\tau$ and $S$,
we can parametrize $F^\tau$ and $F^S$  as follows
\begin{eqnarray}
F^{\tau} =  M_{3/2}^{\prime} (\tau+\overline{\tau}) \cos\theta ~,~\,
F^S = M_{3/2}^{\prime} (S+\overline{S}) \sin\theta~,~\,
\end{eqnarray}
where $ M_{3/2}^{\prime}$ is the gravitino mass if supersymmetry
is only broken by the F-terms of $\tau$ and $S$.
Then, the gaugino masses $M_3$, $M_2$, and $M_1$ respectively
for $SU(3)_C$, $SU(2)_L$, and $U(1)_Y$
at the GUT scale are
\begin{eqnarray}
M_3 &=& {{ \cos\theta + \alpha x \sin\theta} \over\displaystyle
{1 + \alpha x}} M^{\prime}_{3/2} ~,~\, \nonumber \\
M_2 &=&  {{ \cos\theta + (\alpha + 2)x \sin\theta} \over\displaystyle
{1+(\alpha+2)x}} M^{\prime}_{3/2} ~,~\, \nonumber \\
M_1 &=& {{5 \cos\theta + (5\alpha + 6)x \sin\theta} \over\displaystyle
{5+(5\alpha+6)x}} M^{\prime}_{3/2} ~,~\, 
\end{eqnarray}
where $x$ is defined as
\begin{eqnarray}
x &=& {{S+\overline{S}} \over {2(\tau +\overline{\tau})}} ~.~\,
\end{eqnarray}

Using the one-loop renormalization group equations (RGEs), we obtain
the gaugino mass relation around the electroweak scale
\begin{eqnarray}
{{M_1}\over {\alpha_1}} - {{M_3}\over {\alpha_3}}
&=& {3\over 5} \left({{M_2}\over {\alpha_2}} - {{M_3}\over {\alpha_3}} \right)  ~,~\,
\label{GMRelation}
\end{eqnarray}
where $\alpha_3$, $\alpha_2$, and $\alpha_1$ are the gauge couplings for
the $SU(3)_C$, $SU(2)_L$, and $U(1)_Y$ gauge symmetries, respectively.
Following Refs.~\cite{Li:2010xr, Li:2010hi}, we define the index $k$ 
for the gaugino mass relation
as follows
\begin{eqnarray}
k &=& {{{M_2} {\alpha^{-1}_2} - {{M_3} {\alpha^{-1}_3}}}\over
{{{M_1} {\alpha^{-1}_1}} - {{M_3} {\alpha^{-1}_3}}}} ~.~
\end{eqnarray}
Thus, in the F-theory $SU(5)$  and $SO(10)$ models respectively
with $U(1)_Y$ and $U(1)_{B-L}$ fluxes, we obtain that
the index for gaugino mass relation is $5/3$, {\it i.e.},  $k=5/3$.
Moreover, in the minimal supersymmetric Standard Model with
anomaly mediation and mirage mediation~\cite{Choi:2007ka}, we can show that
the index for gaugino mass relation is $5/12$, {\it i.e.},  $k=5/12$~\cite{Li:2010hi}.
Thus, we emphasize that our gaugino mass relation is
different from those in simple anomaly mediation and
mirage mediation~\cite{Choi:2007ka}. 
Furthermore, the index for gaugino mass relation in the minimal
Supergravity (mSUGRA)~\cite{Choi:2007ka} is not well defined 
but can be formally written 
as $0/0$, {\it i.e.},  $k=0/0$~\cite{Li:2010xr}. 
So the gaugino mass relation in mSUGRA 
satisfies the above gaugino mass relation in 
Eq. (\ref{GMRelation}). However,
if $2(M_1\alpha_1^{-1}-M_3 \alpha_3^{-1})/(M_1\alpha_1^{-1}+M_3 \alpha_3^{-1})$ 
is not very small, 
our gaugino mass relation can definitely be distinguished from that
of mSUGRA. Moreover, the gaugino masses can be measured at
the LHC and ILC~\cite{Cho:2007qv, Barger:1999tn, Altunkaynak:2009tg}. Therefore, 
these F-theory GUTs can be tested at the LHC and ILC, and may be
distinguished from the mSUGRA, simple anomaly mediation and mirage 
mediation~\cite{Choi:2007ka}.

To test the gaugino mass relation close to the electroweak scale, we define a
parameter $\eta$ as follows
\begin{eqnarray}
\eta &=& {{ 5\left({{M_1} {\alpha^{-1}_1}} - {{M_3} {\alpha^{-1}_3}}\right)}
\over
 { 3\left({{M_2} {\alpha^{-1}_2}} - {{M_3} {\alpha^{-1}_3}} \right)}}  ~.~\,
\label{Eq-eta}
\end{eqnarray}
Notice $\eta$ is exactly one at the GUT scale. In addition, $\eta$ is one 
around the electroweak scale from one-loop RGE running, yet $\eta$ may deviate slightly
from one as a result of two-loop RGE running.

For simplicity, we assume that $x$ is small in this work, and then
we have approximate gauge coupling unification
at the GUT scale, allowing us to use well-established public codes 
for computations. For gaugino masses, we consider two typical
scenarios

(I) We consider the dilaton dominated scenario, {\it i.e.},
$\theta=\pi/2$. The gaugino
masses at the GUT scale are
\begin{eqnarray}
M_3 & \simeq & \alpha  M_{1/2}  ~,~\, 
M_2 ~ \simeq ~ \left( \alpha  + 2 \right) M_{1/2} ~,~\, \nonumber \\
M_1 & \simeq & \left( \alpha  + {6\over 5}  \right) M_{1/2} ~,~\, 
\end{eqnarray}
where $M_{1/2}$ is a mass parameter. In our numerical calculations, 
we will choose $\alpha=3$.

(II) We consider the scenario where 
$\cos\theta$ is on the order of $ x \sin\theta$.
Assuming $\cos\theta > 0$ and $\sin\theta < 0$, 
we parametrize 
$\cos\theta$ as follows
\begin{eqnarray}
\cos\theta & = & - \gamma x \sin\theta  ~,~\,
\end{eqnarray}
where $\gamma $ is a positive real number.
Thus, we obtain the gaugino
masses at the GUT scale
\begin{eqnarray}
M_3 & \simeq & \left( \gamma - \alpha \right)  M_{1/2}  ~,~\, 
M_2 ~ \simeq ~ \left( \gamma - \alpha  - 2 \right) M_{1/2} ~,~\, \nonumber \\
M_1 & \simeq & \left( \gamma - \alpha  - {6\over 5} \right) M_{1/2} ~.~\,
\end{eqnarray}
In our numerical calculations, 
we choose $(\gamma - \alpha)= 5$. In summary, 
we have $M_3 < M_1 < M_2 $ in scenario I and $M_2 < M_1 < M_3$ in scenario II.

\section{Low Energy Supersymmetry Phenomenology}

We take $\mu > 0$,
so we have four free parameters in our models: 
$M_{1/2}$, $m_0$, $A_{0}$, and $\tan\beta$, where
$\tan\beta$ is the ratio of the Higgs vacuum expectation values.
The soft supersymmetry breaking terms are input into 
{\tt MicrOMEGAs 2.0.7}~\cite{Belanger:2006is} 
using {\tt SuSpect 2.34}~\cite{Djouadi:2002ze} as a front end to 
run the soft terms down to the electroweak scale via RGEs 
and then to calculate the corresponding neutralino relic 
density. We use a top quark mass 
of $m_{t}$ = 173.1 GeV~\cite{:2009ec}. The direct detection cross-sections are 
calculated using {\tt MicrOMEGAs 2.1}~\cite{Belanger:2008sj}. We employ the 
following experimental constraints: (1) The WMAP $2\sigma$ measurements of 
the cold dark matter density~\cite{Spergel:2006hy}, 
0.095 $\leq \Omega_{\chi} \leq$ 0.129. Also, we allow $\Omega_{\chi}$ to be
larger than the upper bound due to a possible $\cal{O}$(10) dilution factor~\cite{Mavromatos:2009pm}
and to be smaller than the lower bound due to multicomponent
dark matter. (2) The experimental limits on 
the Flavor Changing Neutral Current (FCNC) process, $b \rightarrow s\gamma$. 
The results from the Heavy Flavor Averaging Group (HFAG)~\cite{Barberio:2007cr}, 
in addition to the BABAR, Belle, and CLEO results, are: 
$Br(b \rightarrow s\gamma) = (355 \pm 24^{+9}_{-10} \pm 3) \times 10^{-6}$. 
There is also a more recent estimate~\cite{Misiak:2006zs} 
of $Br(b \rightarrow s\gamma) = (3.15 \pm 0.23) \times 10^{-4}$. For our analysis, 
we use the limits 
$2.86 \times 10^{-4} \leq Br(b \rightarrow s\gamma) \leq 4.18 \times 10^{-4}$, 
where experimental and theoretical errors are added in quadrature. 
(3) The anomalous magnetic moment of the muon, $g_{\mu} - 2$. 
For this analysis we use the $2\sigma$ level boundaries, 
$11 \times 10^{-10} < \Delta a_{\mu} < 44 \times 10^{-10}$~\cite{Bennett:2004pv}. 
(4) The process $B_{s}^{0} \rightarrow \mu^+ \mu^-$ where the decay 
has a $\mbox{tan}^6\beta$ dependence. We take the upper bound to be
 $Br(B_{s}^{0} \rightarrow \mu^{+}\mu^{-}) < 5.8 \times 10^{-8}$~\cite{:2007kv}. 
(5) The LEP limit on the lightest CP-even Higgs boson 
mass, $m_{h} \geq 114$ GeV~\cite{Barate:2003sz}.

\begin{figure}[ht]
	\centering
		\includegraphics[width=0.4\textwidth]{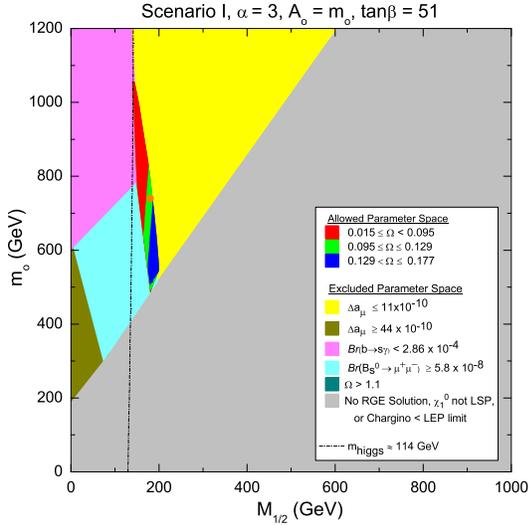}
		\caption{Experimentally allowed parameter space for Scenario I, 
$\alpha = 3,~\beta = 2$, $A_{0} = m_{0}$, tan$\beta$ = 51. The benchmark point detailed 
in Table~\ref{tab:Scenario1_masses} is annotated on the plot by the orange point.}
	\label{fig:mo_scenario1}
\end{figure}

\begin{figure}[ht]
	\centering
		\includegraphics[width=0.4\textwidth]{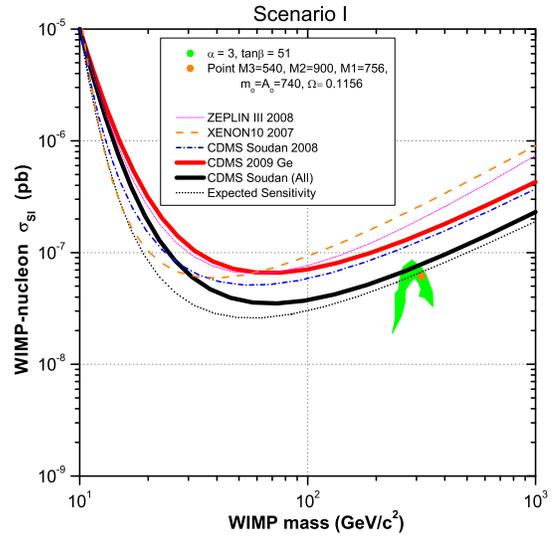}
		\caption{Spin-independent WIMP-nucleon cross-sections for Scenario I, 
$\alpha = 3,~\beta = 2$, $A_{0} = m_{0}$, tan$\beta$ = 51. The green shaded region 
satisfies all experimental constraints. The point detailed 
in Table~\ref{tab:Scenario1_masses} is annotated on the plot by the orange point.}
	\label{fig:cdms_scenario1}
\end{figure}

For scenario I, we commence with $m_{0}$, $A_{0}$, and tan$\beta$ as free parameters, 
however, a comprehensive scan uncovers $A_{0} = m_{0}$ as the most phenomenologically 
favored. As shown in Fig.~\ref{fig:mo_scenario1}, the experimentally allowed parameter 
space for the Scenario I with $\alpha$ = 3, $\beta$ = 2, and tan$\beta$ = 51 after
 applying all these constraints consists of small $M_{1/2}$ and large $m_{0}$. 
We choose a point within the narrow region that satisfies the WMAP relic density
 as our benchmark point for analysis. See table~\ref{tab:Scenario1_masses} for the 
supersymmetric particle (sparticle) and Higgs spectrum. In fact, with a relic density 
of $\Omega_{\chi}$ = 0.1156, 
this benchmark point additionally satisfies the very constrained WMAP 5-year
 results~\cite{Spergel:2006hy}. For constant $m_{0} = A_{0}$ = 740 GeV, 
we find tan$\beta$ = 25-52 for $0 \leq \Omega_{\chi} \lesssim 1.1$, in contrast 
to tan$\beta$ = 41 and tan$\beta$ = 51-52 for the WMAP region. 
In mSUGRA, the focus point region consists of large $m_{0}$ where the WMAP observed 
relic density can be satisfied with a large Higgsino component in the lightest
 supersymmetric particle (LSP) neutralino due to a small $|\mu|$. However, even 
though $m_{0}$ is reasonably large in comparison to $M_{1/2}$ for this benchmark 
point, here the LSP is 98\% bino. The WIMP-nucleon direct-detection 
cross-sections $\sigma_{SI}$ depicted in Fig.~\ref{fig:cdms_scenario1} underscore the fact 
that the case of $\alpha$ = 3, $\beta$ = 2, and tan$\beta$ = 51 produces WIMPs 
with $\sigma_{SI}$ just beneath the CDMS II~\cite{Ahmed:2009zw} upper limit, with 
our benchmark point at $\sigma_{SI} = 6.15 \times 10^{-8}$ pb and 
$m_{\widetilde{\chi}_{1}^{0}}$ = 316 GeV. The constraints from previous 
ZEPLIN~\cite{Lebedenko:2008gb}, XENON~\cite{Angle:2007uj}, and CDMS~\cite{Ahmed:2008eu} 
experiments are also delineated on the plot.

\begin{table}[ht]
  \small
	\centering
	\caption{Supserysmmetric particle (sparticle)
 and Higgs spectrum for a Scenario I, $\alpha = 3,~\beta = 2$ benchmark 
point with $\sigma_{SI} = 6.15 \times 10^{-8}$ pb. Here, tan$\beta$ = 51 and 
$\Omega_{\chi}$ = 0.1156. The GUT scale mass parameters for this point are (in Gev)
 $M_{1/2}$ = 180, $M_3$ = 540, $M_2$ = 900, $M_1$ = 756, $m_{0} = A_{0}$ = 740.}
		\begin{tabular}{|c|c||c|c||c|c||c|c||c|c||c|c|} \hline		
    $\widetilde{\chi}_{1}^{0}$&$316$&$\widetilde{\chi}_{1}^{\pm}$&$473$&$\widetilde{e}_{R}$&$790$&$\widetilde{t}_{1}$&$973$&$\widetilde{u}_{R}$&$1302$&$m_{h}$&$115.4$\\ \hline
    $\widetilde{\chi}_{2}^{0}$&$477$&$\widetilde{\chi}_{2}^{\pm}$&$743$&$\widetilde{e}_{L}$&$946$&$\widetilde{t}_{2}$&$1201$&$\widetilde{u}_{L}$&$1402$&$m_{A}$&$465$\\ \hline
    
    $\widetilde{\chi}_{3}^{0}$&$487$&$\widetilde{\nu}_{e/\mu}$&$942$&$\widetilde{\tau}_{1}$&$489$&$\widetilde{b}_{1}$&$1103$&$\widetilde{d}_{R}$&$1294$&$m_{H^{\pm}}$&$473$\\ \hline
    $\widetilde{\chi}_{4}^{0}$&$743$&$\widetilde{\nu}_{\tau}$&$837$&$\widetilde{\tau}_{2}$&$843$&$\widetilde{b}_{2}$&$1195$&$\widetilde{d}_{L}$&$1404$&$\widetilde{g}$&$1263$\\ \hline

		\end{tabular}
		\label{tab:Scenario1_masses}
\end{table}

\begin{figure}[ht]
	\centering
		\includegraphics[width=0.4\textwidth]{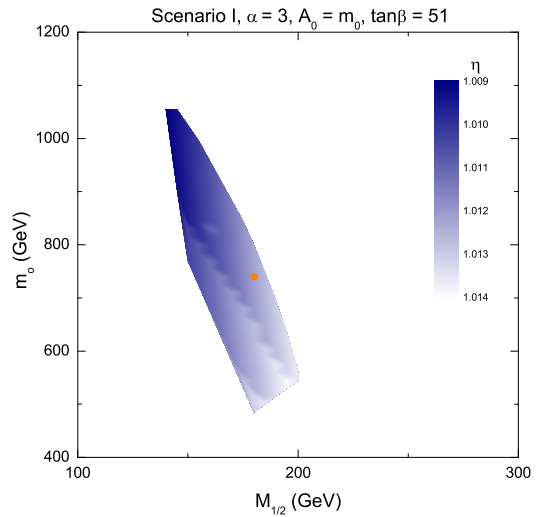}
		\caption{Plot of $\eta$ for Scenario I, 
$\alpha = 3,~\beta = 2$, $A_{0} = m_{0}$, tan$\beta$ = 51. The shaded regions satisfy 
all experimental constraints. The point detailed in Table~\ref{tab:Scenario1_masses} 
is annotated on the plot by the orange point.}
	\label{fig:eta_scenario1}
\end{figure}

The values of $M_1, ~M_2, ~M_3, ~\alpha_{1}, ~\alpha_{2}$ and $\alpha_{3}$ at 
electroweak-symmetry breaking (EWSB) are used to compute $\eta$ in Eq.~(\ref{Eq-eta}), 
and we find the deviation of $\eta$ from one is very small, or on the order 
of 1.2\% - 1.6\%, as expected. The small deviation from one for $\alpha$ = 3,
 $\beta$ = 2, and tan$\beta$ = 51 is clearly shown in Fig.~\ref{fig:eta_scenario1}, 
thus,  the gaugino mass relation in Eq. (\ref{GMRelation}) can be tested
at the LHC and ILC since the two-loop corrections are indeed very small.

The five lightest sparticles for the $\alpha$ = 3, $\beta$ = 2, and tan$\beta$ = 51 
benchmark point, including the heavy Higgs, are 
$\widetilde{\chi}_{1}^{0} < A < H^{\pm} < \widetilde{\chi}_{1}^{\pm} < \widetilde{\chi}_{2}^{0}$. 
The productions of squarks $\widetilde{q}$ and gluino $\widetilde{g}$ have 
the largest differential cross-sections at LHC. The squark in the first two families
will decay dominantly into the gluino $\widetilde{g}$ and the corresponding quark, and
the gluino $\widetilde{g}$ decay will mainly produce either the sbottom $\widetilde{b}$ and bottom quark $b$
or the stop $\widetilde{t}$ and top quark $t$. The $\widetilde{b}$ and
 $\widetilde{t}$ decay to the top quark $t$, bottom quark $b$, neutralino, and 
chargino. Additionally, we will get $b$ quark from $\widetilde{\chi}_{2}^{0}$ through 
light Higgs in the process $\widetilde{\chi}_{2}^{0} \rightarrow h_{0} \widetilde{\chi}_{1}^{0}$ 
with a branching ratio of 93\%. These light Higgs will in turn decay to $b \overline{b}$ 
with a 73\% branching ratio. Leptons and hadronic jets will result from the 
decay $\widetilde{\chi}_{1}^{\pm} \rightarrow W^{\pm} \widetilde{\chi}_{1}^{0}$, where 
this is the only kinematically allowed $\widetilde{\chi}_{1}^{\pm}$ process. Due to 
the less massive nature of the heavy Higgs particles, the $q \overline{q}$ reaction 
will provide some neutral heavy Higgs field $A$ that will decay to $b \overline{b}$ 
pair 87\% of the time, while the heavy charged Higgs field $H^{\pm}$ will 
produce $\overline{b}t$ or $b\overline{t}$ pair 84\% of the time, 
with $t \rightarrow W^{+}b$. Thus, this benchmark point will produce mainly 
light Higgs field $h_{0}$, $b$ quark, and $W$ boson at LHC.

\begin{figure}[ht]
	\centering
		\includegraphics[width=0.4\textwidth]{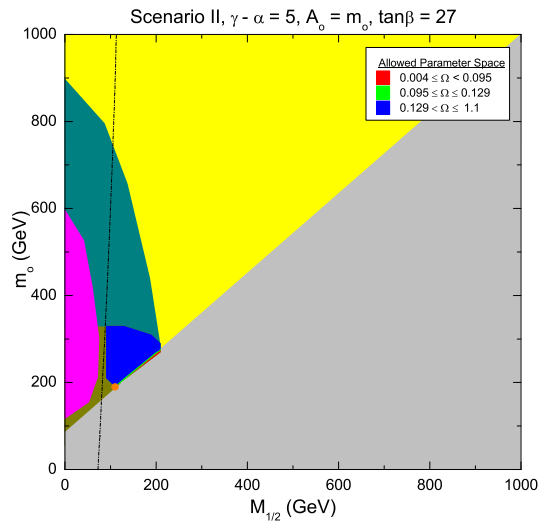}
		\caption{Experimentally allowed parameter space for Scenario II, 
$\gamma - \alpha = 5,~\beta = 2$, $A_{0} = m_{0}$, tan$\beta$ = 27. The benchmark point 
detailed in Table~\ref{tab:Scenario2_masses} is annotated on the plot by the orange point. 
Identification of the excluded regions is shown in the chart legend in Fig.~\ref{fig:mo_scenario1}.}
	\label{fig:mo_scenario2}
\end{figure}

\begin{figure}[ht]
	\centering
		\includegraphics[width=0.4\textwidth]{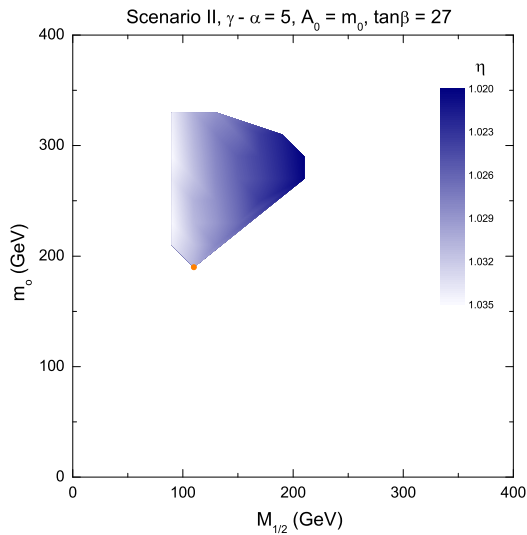}
		\caption{Plot of $\eta$ for Scenario II, 
$\gamma - \alpha = 5,~\beta = 2$, $A_{0} = m_{0}$, tan$\beta$ = 27. The shaded regions 
satisfy all experimental constraints. The point detailed in Table~\ref{tab:Scenario2_masses} is 
annotated on the plot by the orange point.}
	\label{fig:eta_scenario2}
\end{figure}

In addition to the viable parameter space of Scenario I, Scenario II can also generate 
a well constrained region at the relic densities we consider here. Exhibited in 
Fig.~\ref{fig:mo_scenario2} is the experimentally allowed region of 
small $m_{0}$ and $M_{1/2}$ for Scenario I with
 $\gamma - \alpha = 5,~\beta = 2$, $A_{0} = m_{0}$, 
and tan$\beta$ = 27. Choosing a benchmark point with
 $M_{1/2}=110$ GeV and $m_0=A_0=190$ GeV, we present the sparticle and Higgs spectrum 
in Table~\ref{tab:Scenario2_masses}. Because the small mass difference of 10 GeV between 
the LSP neutralino $\widetilde{\chi}_{1}^{0}$ 
and the next to the lightest supersymmetric particle (NLSP) $\widetilde{\tau}_{1}^{\pm}$,
the LSP neutralino in the early Universe can annihilate with stau and then we can
obtain the correct dark matter density,  similar to  
the stau-neutralino coannihilation region in mSUGRA. 
 Moreover, the LSP for this point is 99.7\% bino. 
Considering a constant $m_{0} = A_{0}$ = 190 GeV, we discover tan$\beta$ = 8-59 
for $0 \leq \Omega_{\chi} \lesssim 1.1$, but only tan$\beta$ = 18-27 for the WMAP region. 
This Scenario II benchmark point possesses a $\sigma_{SI} = 2.03 \times 10^{-8}$ pb 
at $m_{\widetilde{\chi}_{1}^{0}}$ = 170 GeV, near the CDMSII upper limit. Furthermore, 
a calculation of $\eta$ in Eq.~(\ref{Eq-eta}) for this Scenario II benchmark point yields 
comparable results to the Scenario I benchmark point, namely only a very small deviation 
from one, on the order of 1.5\% to 3.5\%, as depicted in Fig.~\ref{fig:eta_scenario2}, 
corroborating the delineation of $\eta$ in Fig.~\ref{fig:eta_scenario1} and
 its testability. A close examination of this benchmark point
 reveals $\widetilde{\chi}_{2}^{0} \rightarrow \widetilde{\tau}_{1}^{\mp} 
\tau^{\pm} \rightarrow \tau^{\mp} \tau^{\pm} \widetilde{\chi}_{1}^{0}$ as the dominant decay, 
therefore, we would expect opposite sign tau pair to be characteristic  at the LHC.

\begin{table}[ht]
  \small
	\centering
	\caption{Sparticle and Higgs spectrum for a Scenario II, $\gamma - \alpha = 5,~\beta = 2$ 
benchmark point with $\sigma_{SI} = 2.03 \times 10^{-8}$ pb. Here, tan$\beta$ = 27 and 
$\Omega_{\chi}$ = 0.107. The GUT scale mass parameters for this point are (in Gev)
 $M_{1/2}$ = 110, $M_3$ = 550, $M_2$ = 330, $M_1$ = 418, $m_{0} = A_{0}$ = 190.}
		\begin{tabular}{|c|c||c|c||c|c||c|c||c|c||c|c|} \hline		
    $\widetilde{\chi}_{1}^{0}$&$170$&$\widetilde{\chi}_{1}^{\pm}$&$256$&$\widetilde{e}_{R}$&$247$&$\widetilde{t}_{1}$&$922$&$\widetilde{u}_{R}$&$1117$&$m_{h}$&$115.7$\\ \hline
    $\widetilde{\chi}_{2}^{0}$&$256$&$\widetilde{\chi}_{2}^{\pm}$&$687$&$\widetilde{e}_{L}$&$293$&$\widetilde{t}_{2}$&$1080$&$\widetilde{u}_{L}$&$1126$&$m_{A}$&$661$\\ \hline
    
    $\widetilde{\chi}_{3}^{0}$&$681$&$\widetilde{\nu}_{e/\mu}$&$283$&$\widetilde{\tau}_{1}$&$180$&$\widetilde{b}_{1}$&$1033$&$\widetilde{d}_{R}$&$1115$&$m_{H^{\pm}}$&$666$\\ \hline
    $\widetilde{\chi}_{4}^{0}$&$686$&$\widetilde{\nu}_{\tau}$&$276$&$\widetilde{\tau}_{2}$&$321$&$\widetilde{b}_{2}$&$1095$&$\widetilde{d}_{L}$&$1129$&$\widetilde{g}$&$1257$\\ \hline

		\end{tabular}
		\label{tab:Scenario2_masses}
\end{table}

\begin{figure}[ht]
	\centering
		\includegraphics[width=0.4\textwidth]{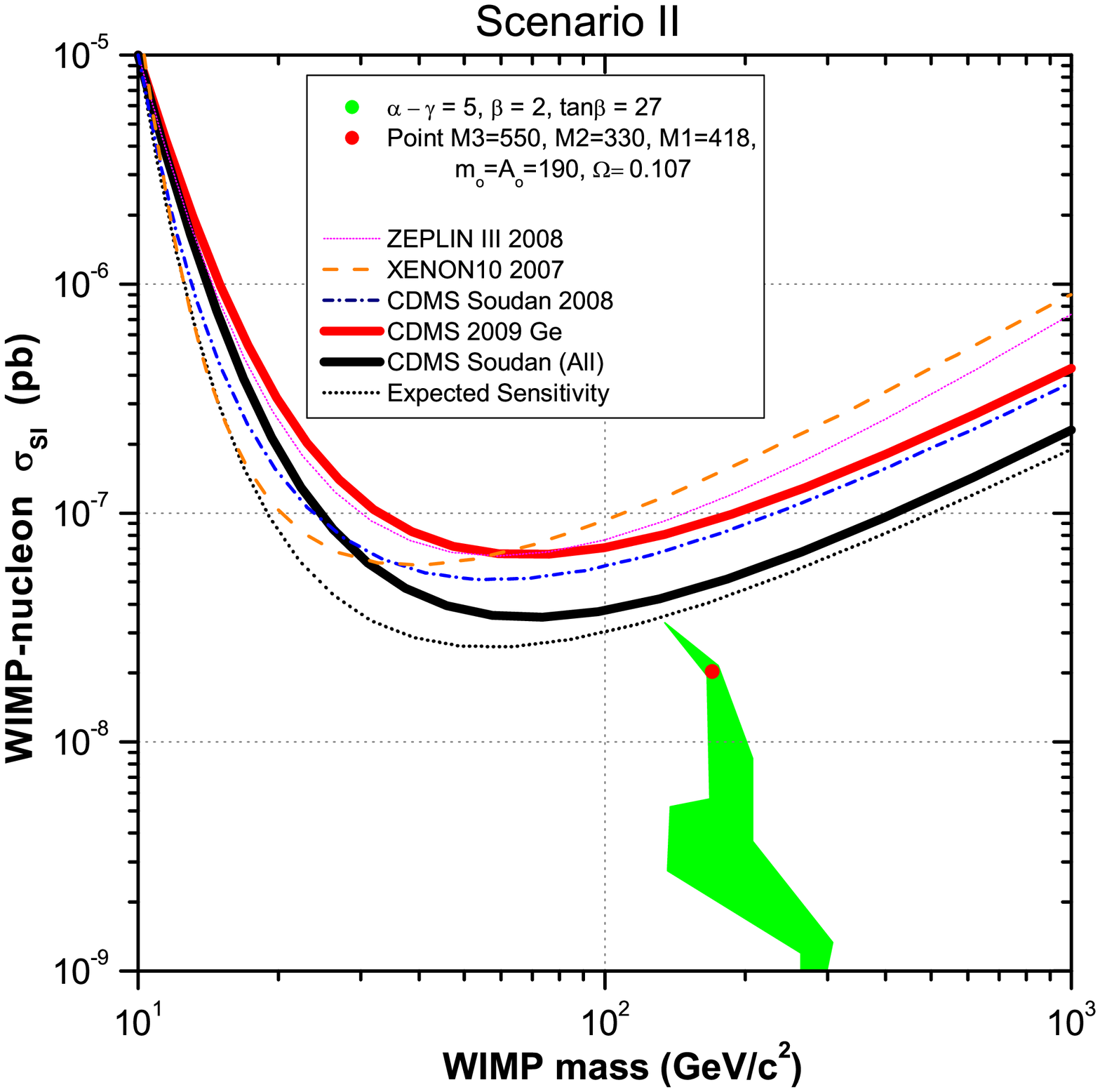}
		\caption{Spin-independent WIMP-nucleon cross-sections for Scenario II, 
$\gamma - \alpha = 5,~\beta = 2$, $A_{0} = m_{0}$, tan$\beta$ = 27. The green shaded region 
satisfies all experimental constraints. The point detailed in Table~\ref{tab:Scenario2_masses} 
is annotated on the plot by the orange point.}
	\label{fig:cdms_scenario2}
\end{figure}

Let us comment on the phenomenological differences between our models and the other
supersymmetry breaking mediation models. Because
 our gaugino mass relation  is different from those in the 
mSUGRA, simple anomaly mediation and mirage mediation, our gaugino mass ratio
$M_1 : M_2 : M_3$ at the low energy are different from those in the
mSUGRA, simple anomaly mediation and mirage mediation. And then the neutralino masses,
chargino masses and gluino mass will be different as well, which will affect
the dark matter density and the productions and decays of the 
supersymmetric particles at the LHC. 

%%%%%%%%%%%%%%%%%%%%%%%%%%%%%%%%%%%%%%%%%%%%%%%%%%%%%%%%%%%%%%%%%%%%%%%%%%%%

%%%%%%%%%%%%%%%%%%%%%%%%%%%%%%%%%%%%%%%%%%%%%%%%%%%%%%%%%%%%%%%%%%%%%%%%%%%%

\section{Conclusion}

We considered gravity mediated 
supersymmetry breaking and
derived the exact gaugino mass relation at one loop
near the electroweak scale in the F-theory  
$SU(5)$ and $SO(10)$ models with $U(1)_Y$ and $U(1)_{B-L}$
fluxes, respectively. The gaugino mass relation presented in this work
differs from the typical gaugino mass relations that
have been studied in the past, and should be preserved
pretty well at low energy in general. Thus,
these F-theory GUTs can be tested at  the LHC
and forthcoming ILC. We exhibited two concrete scenarios 
that satisfy all the latest 
experimental constraints and are consistent with the
CDMS II experiment. Most importantly, the gaugino
mass relation is indeed satisfied at two-loop level
with only a very small deviation.

\section*{Acknowledgments}

%%%%%%%%%%%%%%%%%%%%%%%%%%%%%%%%%%%%%%%%%%%%%%%%%%%%%%%%%%%%%%%%%%%%%%%%%%%%

\begin{acknowledgments}

This research was supported in part 
by  the DOE grant DE-FG03-95-Er-40917 (TL and DVN),
by the Natural Science Foundation of China 
under grant numbers 10821504 and 11075194 (TL),
and by the Mitchell-Heep Chair in High Energy Physics (JM).

\end{acknowledgments}

%%%%%%%%%%%%%%%%%%%%%%%%%%%%%%%%%%%%%%%%%%%%%%%%%%%%%%%%%%%%%%%%%%%%%%%%%%%%

%%%%%%%%%%%%%%%%%%%%%%%%%%%%%%%%%%%%%%%%%%%%%%%%%%%%%%%%%%%%%%%%%%%%%%%%%%%%

\end{document}